\documentclass[11pt]{article}

\usepackage{jheppub} 
\usepackage[english]{babel}
\usepackage{amsmath}

\newcommand{\be}{\begin{equation}}   
	\newcommand{\ee}{\end{equation}}

\preprint{Imperial-TP-2026-CH-1}

\title{Self-Dual Gauge Fields from   Superstring Field Theory}

\author{Chris Hull}
 
\affiliation{The Blackett Laboratory, Imperial College London, Prince Consort Road, London, SW7 2AZ, UK}

\emailAdd{c.hull@imperial.ac.uk}

\abstract
{The action for self-dual gauge fields that emerges from the recently constructed  superstring field theory is found.
The new superstring field theory reduces to that of Sen in a certain limit, and in this limit the new action for self-dual gauge fields reduces to Sen's action for such fields.
 The theory describes two decoupled self-dual gauge fields that couple to two metrics, with each gauge field coupling to only one of the metrics. The action also features a background metric, and the non-linear coupling to these three metrics is non-standard. There are two spin-two gauge invariances, and diffeomorphisms arise from the diagonal subgroup.
  }

\begin{document}
\maketitle

\flushbottom
 
\section{Introduction}

Finding an action for self-dual gauge fields is a long-standing problem; see
\cite{Evnin:2022kqn} for a recent review with an extensive list of references. In \cite{Sen:2015nph,Sen:2019qit}, Sen gave a
remarkable action for a self-dual gauge field that was inspired by his type II
superstring field theory (SFT) \cite{Sen:2015uaa,Sen:2017szq}. The massless modes of the IIB superstring
include a 4-form gauge field with self-dual field strength and any IIB
superstring field theory action must give an action for this self-dual gauge
field. The action for a self-dual 4-form gauge field in 10 dimensions arising
from Sen's SFT can, in fact,  be applied to a self-dual $2 n$-form gauge field in $4 n +
2$ dimensions.

The aim of this paper is to find the self-dual gauge field action that arises
from the new superstring field theory action of \cite{Hull:2025mtb} which, unlike that of \cite{Sen:2015uaa,Sen:2017szq},
is background independent and fully coordinate independent.

Both SFTs are constructed using two string fields, $\Psi, \tilde{\Psi}$ and
both theories have two sectors. One is the physical sector which is the usual
interacting type II superstring with a coupling constant $\kappa$. The other
is a shadow sector which has the same type II superstring spectrum, but is a
free theory in Sen's construction \cite{Sen:2015uaa,Sen:2017szq}, while it is an interacting theory with
coupling constant $\hat{\kappa}$ in the new SFT, with Sen's theory recovered in the limit $\hat \kappa \to 0$. Thus both sectors have the
same spectrum. In both theories, the action has interactions between the two
string fields, but the physical superstring decouples from the shadow
superstring in the field equations. The action can be used for type IIA, IIB
and heterotic superstrings, but here the main focus will be on the IIB
superstring.

Sen's superstring field theory is constructed about a background CFT which
involves a background metric $\bar{g}$. The physical string field $\Psi$ gives
a perturbative graviton $h$ and the physical spacetime metric is then
\begin{equation}
  g_{\mu \nu} = \overline{g }_{\mu \nu} + \kappa h_{\mu \nu} \label{fism} .
\end{equation}
The shadow sector includes a second graviton $\hat{h}$, which is a free
spin-two field coupled to the background $\bar{g}$. 
For the case in which the background spacetime  is 10-dimensional Minkowski space
(with  $\bar{g}$ the Minkowski metric), it was shown in \cite{Mamade:2025jbs} that, at least to cubic order in the fields,  the IIB superstring field
theory gives a physical self-dual 4-form gauge field coupling to $g$, together
with a shadow sector self-dual 4-form gauge field coupling to the Minkowski metric $\bar{g}$. Moreover, it was shown that the action that is in agreement with Sen's action for these self-dual gauge fields (to cubic order in the fields).
For general backgrounds, it is then to be expected that the IIB SFT gives a physical self-dual 4-form gauge field coupling to $g$, together
with a shadow sector self-dual 4-form gauge field coupling to $\bar{g}$.

In contrast, for the superstring field theory of \cite{Hull:2025mtb}, both the physical
and the shadow superstrings are fully interacting. The second graviton
$\hat{h}$ arising in the shadow superstring is then self-interacting and
couples to all other fields in the shadow sector. The shadow theory couples to
the metric
\begin{equation}
  \hat{g}_{\mu \nu} = \bar{g}_{\mu \nu} + \hat{\kappa} \hat{h}_{\mu \nu},
\end{equation}
while the physical sector again couples to the physical metric (\ref{fism}).
The IIB superstring field theory gives a physical self-dual gauge field
coupling to $g$, together with a shadow sector self-dual gauge field now
coupling to $\hat{g}$.

In this paper, the actions for self-dual gauge fields of \cite{Sen:2015nph,Sen:2019qit} and \cite{Hull:2023dgp}
will be modified to give the action that should arise (for self-dual 4-forms
in 10 dimensions) from the new superstring field theory, and check that the
cubic couplings agree with those of the SFT. The action has a physical sector
consisting of the physical gauge field and the graviton $h$ which couple to
the other physical fields, while there is also a shadow sector consisting of
the shadow gauge field and the shadow graviton $\hat{h}$. The shadow sector is
interacting but completely decouples from the physical sector.

The resulting self-dual gauge theory can be applied to self-dual gauge fields
in $4 n + 2$ dimensions and involves three metrics, the background metric
$\overline{\! g}$, the physical metric $g$ and the shadow sector metric
$\hat{g}$. The variables in the action are `primitive' fields from which the
physical and shadow fields are constructed, and these couple to all three
metrics. A non-trivial part of the construction is finding these novel
couplings to three metrics, and draws on the bi-metric geometry developed in
\cite{Hull:2023dgp}.

In section 2, the field equations are given and then in the following sections
the action is constructed. Section 3 reviews the free theory of self-dual gauge fields coupling only to the
background metric $\bar{g} .$ In section 4, the linearised couplings to the
two gravitons is given and shown to agree with the action from SFT.  In
section 5, the bi-metric generalisation of Sen's action is reviewed and in
section 6 a new action is proposed with non-polynomial couplings to the two
metrics $g, \hat{g}$ and it is shown to give the field equations of section 2
and the cubic couplings of section 3. In section 7 the symmetries of the new
action are found and section 8 provides a discussion of the results presented
here.

\section{The field equations}. \label{feqns}

In this section, the field equations of the theories for self-dual gauge
fields in \cite{Sen:2015nph,Sen:2019qit} and \cite{Hull:2023dgp} will be reviewed and field equations for the
new theory will be proposed. In $d = 2 q$ dimensions with $q$ an odd integer,
the degrees of freedom are the physical $q - 1$-form gauge field with
self-dual $q$-form field strength, together with a second gauge field with
self-dual $q$-form field strength. The physical gauge field couples to gravity
and the other physical fields, while the second one is part of a shadow sector
that completely decouples from the physical sector. In particular, it does not
couple to the physical graviton $h$. These features make the coupling to
gravity non-standard.

A useful starting point is the free theory in $2 q$-dimensional Minkowski
space with metric $\eta_{\mu \nu}$ \cite{Hull:2025bqo}, with two $q$-form field strengths
$F_0, G_0$ that are self-dual with respect to the Minkowski space Hodge
duality operator $\ast_{\eta}$:
\begin{equation}
  F_0 = \ast_{\eta} F_0, \quad G_0 = \ast_{\eta} G_0 .
\end{equation}
They are closed
\begin{equation}
  d F_0 = 0 \quad d G_0 = 0,
\end{equation}
and hence co-closed, being annihilated by $\ast_{\eta} d \ast_{\eta}$. This is then
a theory of two self-dual gauge fields in Minkowski space. The action will be
discussed in the next section.

It was argued in \cite{Hull:2025bqo} that Sen's theory can be viewed as the coupling of this
theory to a metric $g$ in such a way that the solutions are now expressed in
terms of $G_0 $ and a field strength $F$ that is self-dual with respect to the
Hodge duality operator $\ast_g$ for the metric $g$:
\begin{equation}
  F  = \ast_g F , \quad G_0 = \ast_{\eta} G_0 \label{etfeq} .
\end{equation}
Both field strengths are closed,
\begin{equation}
  d F  = 0, \quad d G_0 = 0 \label{clo} .
\end{equation}
Then $F$ couples to the physical metric $g$ while $G_0 $ does not couple to $g$
but only couples to $\eta$. This  gives a theory of a physical gauge field
with field strength $F$ that is self-dual with respect to the physical metric
$g$ together with another gauge field whose field strength $G_0 $ is self-dual
with respect to the Minkowski metric. Sen showed that the physical fields $F,
g$ can then be coupled to further physical fields from which $G_0$ remains
decoupled. The action will be discussed in later sections.

The new field strength $F$ is given in terms of the original $F_0$ by
\begin{equation}
  F = F_0 + M (F_0, g, \eta),
\end{equation}
where the function $M$ depends on both $g$ and $\eta$ and this expression
converts the $\eta$-self-dual form $F_0$ to the $g$-self-dual one $F$. Here,
the focus will be on Maxwell-like theories  for which $M$ is a linear
function of $F_0 $, but the theory generalises to Born-Infeld type theories
for which $M$ is non-linear. Sen constructed $M$ perturbatively as a power
series in $h$ where
\begin{equation}
  g_{\mu \nu} = \eta_{\mu \nu} + \kappa h_{\mu \nu} .
\end{equation}
To lowest order in $h$, it follows from \cite{Sen:2015nph,Sen:2019qit,Hull:2023dgp,Andriolo:2020ykk}  that
\begin{equation}
  M (F_0) = \kappa h \cdot F_0 + O (h^2) \label{mfo}
\end{equation}
where for any $q$-form $Q$ we use the notation of \cite{Mamade:2025jbs}
\begin{equation}
  (h \cdot Q)_{\mu_1 \mu_2 \ldots \mu_q} = \eta^{\nu \rho} h_{\nu [\mu_1
  } Q_{ \mu_2 \cdots \mu_q] \rho} .
\end{equation}
A non-perturbative construction of $M$ was given in \cite{Andriolo:2020ykk,Hull:2023dgp}. It was shown
in \cite{Mamade:2025jbs} that Sen's theory arises from Sen's superstring field theory \cite{Sen:2015uaa,Sen:2017szq}, to linear
order in $h$.

This formulation is restricted to spacetimes that admit a Minkowski metric.
In \cite{Hull:2023dgp}, a generalisation was found in which the Minkowski spacetime with
metric $\eta$ is replaced by an arbitrary background spacetime with metric
$\bar{g}$. The free CFT then gives field strengths that remain closed,
(\ref{clo}), and are self-dual with respect to the Hodge duality operator
$\ast_{\bar{g}}$ for $\bar{g}$:
\begin{equation}
  F_0 = \ast_{\bar{g}} F_0, \quad G_0 = \ast_{\bar{g}} G_0 .
\end{equation}
Coupling to a metric $g$ now gives
\begin{equation}
  F  = \ast_g F , \quad G_0 = \ast_{\bar{g}} G_0 \label{gheq} .
\end{equation}
As before, $F$ is expressed as
\begin{equation}
  F = F_0 + M (F_0, g, \bar{g})
\end{equation}
but where now $M$ depends on $F_0, g, \bar{g}$. Writing
\begin{equation}
  g_{\mu \nu} = \bar{g}_{\mu \nu} + \kappa h_{\mu \nu},
\end{equation}
$M$ is again of the form (\ref{mfo}) but now with
\begin{equation}
  (h \cdot Q)_{\mu_1 \mu_2 \ldots \mu_q} = \bar{g}^{\nu \rho} h_{\nu [\mu_1
  } Q_{ \mu_2 \cdots \mu_q] \rho} .
\end{equation}

The generalisation of this that arises from the string field theory of \cite{Hull:2025mtb} introduces a second  graviton $\hat{h}$ (in the shadow sector) which
yields a further metric
\begin{equation}
  \hat{g}_{\mu \nu} = \bar{g}_{\mu \nu} + \hat{\kappa} \hat{h}_{\mu \nu} .
\end{equation}
The theory modifies both field strengths so that
\begin{eqnarray}
  F & = & F_0 + M (F_0, g, \bar{g}) \nonumber\\
  G & = & G_0 + \hat M (G_0, \hat{g}, \bar{g}) 
\end{eqnarray}
with
\begin{eqnarray}
  F & = & \ast_g F \nonumber\\
  G & = & \ast_{\hat{g}} G  \label{geq}
\end{eqnarray}
where $\ast_{\hat{g}}$ is the Hodge dual for the metric $\hat{g}$, and
\begin{equation}
  d F = 0, \quad d G = 0 \label{gclo} .
\end{equation}
Here $\hat M (G_0, \hat{g}, \bar{g})$ has the same functional form as $M (F_0, g,
\bar{g})$, so that
\begin{eqnarray}
  M (F_0, g, \bar{g}) & = & \kappa h \cdot F_0 + O (\kappa^2 h^2), \nonumber\\
 \hat  M (G_0, \hat{g}, \bar{g}) & = & \hat{\kappa} \hat{h} \cdot G_0 + O
  (\hat{\kappa}^2 \hat{h}^2) . 
\end{eqnarray}
This is then a theory of a gauge field with field strength $F$ that is
self-dual with respect to the metric $g$ together with another gauge field
whose field strength $G  $ is self-dual with respect to the metric $\hat{g}$.

\section{Free action}

The generalisation {\cite{Hull:2023dgp}} of Sen's theory
{\cite{Sen:2015nph,Sen:2019qit}} is a theory with two metrics on the
spacetime: a ``physical'' metric $g_{\mu \nu}$ which couples to all the
physical fields and carries the gravitational degrees of freedom, and a second
metric $\bar{g}_{\mu \nu}$ which does not couple to the physical fields. Sen's
theory is recovered on setting the second metric to be the Minkowski metric,
$\bar{g}_{\mu \nu} = \eta_{\mu \nu}$. In this section, the simple conformal
theory of \cite{Hull:2025bqo} obtained by setting the two metrics equal, $\bar{g}_{\mu \nu} =
g_{\mu \nu}$, will be reviewed.

The degrees of freedom of the theory in $d = 2 q$ dimensions with $q = 2 n +
1$ are a $q - 1$-form field $P$ and a $q$-form field $Q$ which is
$\bar{g}$-self-dual with respect to $\bar{g}_{\mu \nu}$, $Q = \bar{\ast} Q$
where $\bar{\ast} = \ast_{\bar{g}}$ is the Hodge dual with respect to the
metric $\bar{g}$. The free action, coupled to the background metric $\bar{g}$,
is
\begin{equation}
  \label{actfree} S = \int \left( Q \wedge dP - \frac{1}{2} dP \wedge
  \bar{\ast} dP \right).
\end{equation}
Following \cite{Hull:2025bqo}, the shift
\begin{equation}
  Q' = Q + \Omega , \label{qpr}
\end{equation}
where
\begin{equation}
  \Omega  = \frac{1}{2}  [dP + \bar{\ast} dP], \label{omis}
\end{equation}
 gives the action
\begin{equation}
  \label{act222} S = \int Q' \wedge dP.
\end{equation}
Note that $Q'$ is self-dual $Q' = \bar{\ast} Q'$.
The field equations are
\begin{equation}
  dQ' = 0
\end{equation}
and
\begin{equation}
  dP = \bar{\ast} dP
\end{equation}
so that $Q'$ and $d P$ are both closed and $\bar{\ast}$-self-dual and hence
co-closed, annihilated by $\bar{\ast} d \bar{\ast}$. Note that, without the
self-duality constraint on $Q'$, this would be a BF theory which is
topological, but here the self-duality constraint leads to non-trivial
dynamics, giving a higher-dimensional generalisation of the $\beta \gamma$
system \cite{Hull:2025bqo}.

Following \cite{Hull:2023dgp,Hull:2025bqo}, it is useful to define the field strengths
\begin{equation}
  \label{Gissa} G_0 = Q + 2\Omega
\end{equation}
and
\begin{equation}
  \label{fisa} F_0 = Q\, .
\end{equation}
In terms of $Q'$, these are:
\begin{equation}
  \label{Gissap} G_0 = Q' + \Omega\, ,
\end{equation}
and
\begin{equation}
  \label{fop} F_0 = Q' - \Omega\, .
\end{equation}
These are both closed and self-dual,
\begin{equation}
  dF_0 = 0, \qquad dG_0 = 0, \qquad F_0 = \bar{\ast} F_0, \qquad G_0 =
  \bar{\ast} G_0 .
\end{equation}
This implies that  there are local potentials $A_0, C_0$ with $F_0 = dA_0$, $G_0 = dC_0$.

\section{Linearised Coupling to Gravitons and Comparison to SFT}

\subsection{Coupling to Gravitons}

The natural energy-momentum tensors for the self-dual fields $F_0$, $G_0$ are
\begin{eqnarray}
  T_{\mu \nu} & = & \frac{1}{4(q - 1) !} \bar{g}^{\rho_1 \lambda_1} \ldots
  \bar{g}^{\rho_{q - 1} \lambda_{q - 1}} (F_0)_{\mu \rho_1 \ldots \rho_{q -
  1}}  (F_0)_{\nu \lambda_1 \ldots \lambda_{q - 1}}  \label{Tmn}\\
  \hat{T}_{\mu \nu} & = & \frac{1}{4(q - 1) !} \bar{g}^{\rho_1 \lambda_1}
  \ldots \bar{g}^{\rho_{q - 1} \lambda_{q - 1}} (G_0)_{\mu \rho_1 \ldots
  \rho_{q - 1}}  (G_0)_{\nu \lambda_1 \ldots \lambda_{q - 1}} .  \label{Tmnh}
\end{eqnarray}

For Sen's theory, the background metric is the Minkowski metric: $\overline{\!
g }_{\mu \nu} = \eta_{\mu \nu}$. In \cite{Hull:2025bqo} it was shown that Sen's theory can be
viewed as first adding to (\ref{actfree}) (with $\bar{g }_{\mu \nu} =
\eta_{\mu \nu}$) the linear coupling to $h$
\begin{equation}
  \frac{\kappa}{2} \int d^{2 q} x \, h_{\mu \nu} T^{\mu \nu} \label{tmn}
\end{equation}
and then adding higher order terms in $h$ to obtain a gauge-invariant action
that is non-polynomial in $h$. The bi-metric action \cite{Hull:2023dgp} can be obtained by a
similar construction with general $\bar{g}$. The full non-linear action will
be reviewed in section \ref{bimac} and shown to give the field equations
(\ref{clo}),(\ref{gheq}).

The new action is obtained by further adding the linear coupling
\begin{equation}
  - \frac{\hat{\kappa}}{2} \int d^{2 q} x\,  \hat{h}_{\mu \nu} \hat{T}^{\mu \nu}
  \label{tmnh}
\end{equation}
and then completing to a non-linear theory that is non-polynomial in both
$\kappa h_{\mu \nu}$ and $\hat{\kappa} \hat{h}_{\mu \nu}$. The full action
will be presented in section \ref{trimact} and shown to give the field
equations (\ref{geq}),(\ref{gclo}). Note that the two couplings
(\ref{tmn}),(\ref{tmnh}) have opposite signs: this is important for the
construction of the non-linear theory that arises here.

\subsection{Self-Dual Gauge Field Actions from 
String Field Theory}

Sen's SFT \cite{Sen:2015uaa,Sen:2017szq} is formulated in terms of two string fields, $\Psi$ and
$\tilde{\Psi}$. In \cite{Mamade:2025jbs} it was shown that the IIB SFT action in a Minkowski
space background gives a RR kinetic term for a self-dual 5-form $Q$ and a
4-form $P$ in 10 dimensions which is of precisely the form (\ref{actfree}).
Here $Q$ originates from the string field $\Psi$ while $P$ comes from the
second string field $\tilde{\Psi}$.

The expansion of Sen's SFT action in 10-dimensional Minkowski space to cubic
order in fields was calculated in \cite{Mamade:2025jbs}. There it was shown that the term
quadratic in the massless RR gauge fields and linear in the NS-NS graviton $h$
is precisely (\ref{tmn}). A similar expansion of the new SFT \cite{Hull:2025mtb} gives a term
quadratic in the massless RR gauge fields and linear in the gravitons $h,
\hat{h}$ is precisely the sum of (\ref{tmn}) and (\ref{tmnh}). The calculation
of the cubic terms involving $\hat{h}$ is very similar to that of the terms
involving $h$.

\section{The Bi-metric Generalisation of Sen's Action}\label{bimac}

The generalisation {\cite{Hull:2023dgp}} of Sen's theory
{\cite{Sen:2015nph,Sen:2019qit}}  with two metrics $g_{\mu \nu}$ and
$\bar{g}_{\mu \nu}$ has the action\footnote{The $Q,F,G$ here have been rescaled compared to those in  \cite{Andriolo:2020ykk,Hull:2023dgp}. The $Q,F,G$  here are related to the $Q_{\rm {old}}$, $F_{\rm {old}},G_{\rm {old}}$ in \cite{Andriolo:2020ykk,Hull:2023dgp} by $Q=2Q_{\rm {old}}$, $F=2F_{\rm {old}}$, $G=2G_{\rm {old}}$. }
\begin{equation}
  S = \int \left( Q \wedge dP - \frac{1}{2} dP \wedge \bar{\ast} dP + V (Q)
  \right), \label{2act}
\end{equation}
where
\begin{equation}
  V (Q) = \frac{1}{4} Q \wedge M (Q) .
\end{equation}
Here $M (Q)$ is a function of $Q$ depending on both metrics, $g_{\mu \nu}$ and
$\bar{g}_{\mu \nu}$, and is given by a particular function $\mathcal{M} (Q,
\bar{g}_{\mu \nu}, g_{\mu \nu})$ that was constructed in \cite{Andriolo:2020ykk,Hull:2023dgp} so that
\begin{equation}
  M (Q) =\mathcal{M} (Q, \bar{g}_{\mu \nu}, g_{\mu \nu}) \label{mcal} .
\end{equation}
The field equations are {\cite{Hull:2023dgp}}
\begin{equation}
  \label{feq1} d (\bar{\ast} dP + Q) = 0,
\end{equation}
\begin{equation}
  \label{feq2} M (Q) + dP - \bar{\ast} dP = 0.
\end{equation}
Then $G_0$ defined by (\ref{Gissa}) is $\bar{g}$-self-dual
\begin{equation}
  \bar{\ast} G_0 = G_0,
\end{equation}
and defining
\begin{equation}
  \label{Fiss} F \equiv F_0 + M (F_0),
\end{equation}
the field equations (\ref{feq1}),(\ref{feq2}) imply
\begin{equation}
  dG_0 = 0, \qquad dF = 0.
\end{equation}
There are then $q - 1$-form potentials $A, C$ with $F = dA$, $G_0 = dC_0$. The
key point is that $M (Q)$ is constructed so that $F$ is $g$-self-dual
\begin{equation}
  \ast F = F.
\end{equation}
so that the field equations (\ref{clo}),(\ref{gheq}) are recovered.

This is then a theory of the desired $(q - 1)$-form $A$ with self-dual field
strength $\ast F = F$ and an auxiliary $(q - 1)$-form $C_0$ whose field
strength is self-dual with respect to the background metric, $\bar{\ast} G_0 =
G_0$. It is important that the auxiliary field $C_0$ does not couple to the
physical metric $g_{\mu \nu}$ and the physical field $A$ does not couple to
the auxiliary metric $\bar{g}_{\mu \nu}$.

The action (\ref{2act}) can be rewritten in terms of $Q', P$ as
\begin{equation}
  \label{act222} S = \int (Q' \wedge dP + V (F_0))
\end{equation}
with $F_0$ defined by (\ref{fop}).

\section{New Action}\label{trimact}

The new action involves couplings to the shadow sector metric $\hat{g}_{\mu
\nu}$. In analogy with (\ref{Fiss}) and following the discussion in section \ref{feqns}, we aim to construct a field strength
\begin{equation}
  G = G_0 + \hat{M} (G_0) \label{Gisz}
\end{equation}
so that adding $\hat{M} (G_0)$ takes the $\bar{g}$-self-dual field strength
$G_0$ to one that is self-dual with respect to $\hat{g}$:
\begin{equation}
  G = \hat{\ast} G.
\end{equation}
Here $\hat{\ast}$ is the Hodge dual for the metric $\hat{g}_{\mu \nu}$. This
can be done by choosing
\begin{equation}
  \hat{M} (G_0) =\mathcal{M} (G_0, \bar{g}_{\mu \nu}, \hat{g}_{\mu \nu})
\end{equation}
where $\mathcal{M}$ is the same function that appears in (\ref{mcal}), but
with $Q = F_0$ replaced by $G_0$ and $g$ replaced by $\hat{g}$.

The new action is
\begin{equation}
  \label{actnew} S = \int \left( Q \wedge dP - \frac{1}{2} dP \wedge
  \bar{\ast} dP + V (F_0) - \hat{V} (G_0) \right),
\end{equation}
where $\hat{V} (Q)$ is given by
\begin{equation}
  \hat{V} (Q) = \frac{1}{4} Q \wedge \hat{M} (Q)
\end{equation}
and $F_0 = Q$ while $G_0$ is given by (\ref{Gissa}). Varying with respect to
$Q$ gives
\begin{equation}
  (d P - \bar{\ast} d P) + M (F_0) - \hat{M} (G_0) = 0 \label{qeq}
\end{equation}
while varying with respect to $P$ gives
\begin{equation}
  d (\bar{\ast} dP + Q + \hat{M} (G_0)) = 0 \label{peq} .
\end{equation}
Taking the exterior derivative of (\ref{qeq}) gives
\begin{equation}
  d (\bar{\ast} dP - M (F_0) + \hat{M} (G_0)) = 0
\end{equation}
and subtracting (\ref{peq}) from this gives
\begin{equation}
  d (Q + M (F_0)) = 0
\end{equation}
which, using (\ref{fisa}),(\ref{Fiss}), gives
\begin{equation}
  d F = 0.
\end{equation}
Furthermore, (\ref{peq}) implies
\begin{equation}
  d (\bar{\ast} dP + d P + Q + \hat{M} (G_0)) = 0
\end{equation}
which, using (\ref{Gissa}),(\ref{Gisz}), gives
\begin{equation}
  d G = 0.
\end{equation}
Thus the result is two closed field strengths $F, G$ with the self-duality
conditions
\begin{equation}
  F = \ast F, \qquad G = \hat{\ast} G
\end{equation}
Thus the desired field equations (\ref{geq}),(\ref{gclo}) are indeed obtained.

The action (\ref{actnew}) can be written in terms of $Q', P$ takes the
symmetric form
\begin{equation}
  \label{act222} S = \int \{ Q' \wedge dP + V (Q' - \Omega) - \hat{V} (Q' +
  \Omega) \},
\end{equation}
and from (\ref{qpr})
\begin{equation}
  Q' = Q + \Omega .
\end{equation}

\section{Symmetries}

\subsection{Bi-metric Action}

The symmetries of the bi-metric action (\ref{2act}) were found in \cite{Hull:2023dgp}.
Firstly, it is invariant under transformations in which $g$ transforms as a
spin-two gauge field but $\bar{g}$ is invariant:
\begin{eqnarray}
  \delta g =\mathcal{L}_{\zeta} g, &  & \delta \bar{g} = 0, \nonumber\\
  \delta P = \frac{1}{2} i_{\zeta} F, &  & \delta Q = - (1 + \bar{\ast}) d
  \delta P  \label{zsymm}
\end{eqnarray}
where $\mathcal{L}_{\zeta}$ is the Lie derivative with respect to $\zeta$.
These imply the invariance of the field strength $G_0$

\begin{equation}
  \label{cdiff} \delta G_0 = 0.
\end{equation}

Following \cite{Andriolo:2020ykk,Hull:2023dgp} it is useful to introduce a map $\Psi = 1 + M$ from
$q$-forms to $q$-forms
\begin{equation}
  \label{siis} \Psi (Y) = Y + M (Y) .
\end{equation}
For any $q$-form $Y$
\begin{equation}
  \Pi_- \Psi (\bar{\Pi}_+ Y) = 0,
\end{equation}
so that $\Psi$ takes a $\bar{g}$-self-dual form $X$ (with $X = \bar{\ast} X$)
to a $g$-self-dual one $\Psi (X)$ (with $\Psi (X) = \ast \Psi (X)$). Then
\begin{equation}
  \delta F =\mathcal{L}_{\zeta} F - \Psi (\bar{\Pi}_+ i_{\zeta} dF) .
\end{equation}
The field equations imply $dF = 0$, so that on-shell
\begin{equation}
  \delta F \approx \mathcal{L}_{\zeta} F.
\end{equation}

Note that (\ref{zsymm}) implies that
\begin{equation}
  \delta Q = - 2 \delta \Omega,
\end{equation}
where $\Omega$ is defined by (\ref{omis}), so that
\begin{equation}
  \delta Q' = - \delta \Omega = - \frac{1}{2} (1 + \bar{\ast}) d \delta P
  \label{qptrans}
\end{equation}
and $Q' + P$ is invariant:
\begin{equation}
  \delta (Q' + \Omega) = 0.
\end{equation}
For $\bar{g} = \eta$, these transformations are the symmetry of Sen's action
discussed in \cite{Andriolo:2020ykk}.

The theory is also invariant under diffeomorphisms in which all fields
transform covariantly
\begin{eqnarray}
  \delta g =\mathcal{L}_{\xi} g, &  & \delta \bar{g} =\mathcal{L}_{\xi} 
  \bar{g} \nonumber\\
  \delta Q =\mathcal{L}_{\xi} Q &  & \delta P =\mathcal{L}_{\xi} P 
  \label{diffs}
\end{eqnarray}
and antisymmetric gauge transformations
\begin{equation}
\label{Pgag}
  \delta P = d \alpha .
\end{equation}

Finally, combining a symmetry (\ref{zsymm}) and a diffeomorphsim (\ref{diffs})
with parameters $\chi \equiv \xi = - \zeta$ gives a symmetry
\begin{eqnarray}
  \delta \bar{g} =\mathcal{L}_{\chi}  \bar{g} &  & \delta g = 0 \nonumber\\
  \delta P =\mathcal{L}_{\chi} P - \frac{1}{2} i_{\chi} F &  & \delta Q
  =\mathcal{L}_{\chi} Q + \frac{1}{2} (1 + \bar{\ast}) d i_{\chi} F 
  \label{chsym}
\end{eqnarray}
under which $g$ is invariant while $\bar{g}$ transforms as a spin-two gauge
field.

\subsection{Tri-metric Action} \label{trisect}

Consider now the new tri-metric action (\ref{actnew}). The theory is again
invariant under diffeomorphisms in which all fields transform covariantly
\begin{eqnarray}
  \delta g =\mathcal{L}_{\xi} g, &  & \delta \bar{g} =\mathcal{L}_{\xi} 
  \bar{g}, \quad \delta \hat{g} =\mathcal{L}_{\xi}  \hat{g}, \nonumber\\
  \delta Q =\mathcal{L}_{\xi} Q &  & \delta P =\mathcal{L}_{\xi} P 
  \label{diffs2}
\end{eqnarray}
and antisymmetric gauge transformations
\begin{equation}
  \delta P = d \alpha .
\end{equation}

Next, consider the symmetry (\ref{zsymm}) for which $g$ is a gauge field and
$\bar{g}$ is invariant. Under the transformations (\ref{zsymm}), $\bar{g},
G_0$ are invariant. If $\hat{g}$ is also taken to be invariant,
\begin{equation}
  \delta \hat{g} = 0 \label{gha},
\end{equation}
then the extra term $\hat{V} (G_0) = \hat{V} (Q' + \Omega)$ is invariant, so
that the whole action (\ref{actnew}) or (\ref{act222}) is invariant. Hence the
action is invariant under (\ref{zsymm}),(\ref{gha}). Note that $G = G_0 +
\hat{M} (G_0)$ is invariant as $G_0$ and $\hat{g} $are. Moreover,
\begin{equation}
  \delta_{\zeta} Q' = - \delta_{\zeta} \Omega = \frac{1}{2} (1 + \bar{\ast}) d
  \delta P \label{qptransas}
\end{equation}
so that
\begin{equation}
  \delta_{\zeta} (Q' + \Omega) = 0.
\end{equation}
Under these transformations
\begin{equation}
  \delta G =0
\end{equation}
and on-shell
\begin{equation}
  \delta F \approx \mathcal{L}_{\zeta} F.
\end{equation}

The action (\ref{act222}) changes by a sign, $S \rightarrow - S$, under
\begin{equation}
  P \rightarrow - P, \quad g \rightarrow \hat{g}, \quad \hat{g} \rightarrow g,
  \quad Q' \rightarrow Q' \label{z2sym} 
\end{equation}
wiht $Q'$ invariant.
From (\ref{Gissa}),(\ref{fop}), this gives
\begin{equation}
  F_0 \rightarrow G_0, \quad G_0 \rightarrow F_0 \label{z2symf} .
\end{equation}
Combining this fact with the invariance under (\ref{zsymm}),(\ref{gha})
implies that the action is also invariant under
\begin{equation}
  \delta Q  = 0, \quad \delta \bar{g} = 0, \quad \delta g = 0, \quad \delta
  \hat{g} =\mathcal{L}_{\hat{\zeta}}  \hat{g} \label{chit},
\end{equation}
together with
\begin{equation}
  \delta P = - \frac{1}{2} i_{\hat{\zeta}} G \label{chitp}
\end{equation}
These then imply that
\begin{equation}
  \delta F = 0, \quad \delta G \approx \mathcal{L}_{\hat{\zeta}} G.
\end{equation}
Moreover,
\begin{equation}
  \delta Q' = \delta_{\hat{\zeta}} \Omega \label{qptransasa}
\end{equation}
so that
\begin{equation}
  \delta_{\hat{\zeta}} (Q' - \Omega) = 0.
\end{equation}
These transformations are obtained from (\ref{zsymm}),(\ref{gha}) by
interchanging $g$ with $\hat{g}$ and $F$ with $G$ while taking $P \rightarrow
- P$ and keeping $Q'$ fixed. This invariance can also be checked directly,
following the analysis in \cite{Hull:2023dgp}.

Under the diagonal subgroup of the $\zeta, \hat{\zeta}$ transformations with
$\zeta = \hat{\zeta} \equiv \hat{\xi}$
\begin{equation}
  \delta g =\mathcal{L}_{\hat{\xi}} g, \quad \delta \hat{g}
  =\mathcal{L}_{\hat{\xi}} \hat{g}, \quad \delta F \approx
  \mathcal{L}_{\hat{\xi}} F, \quad \delta G \approx \mathcal{L}_{\hat{\xi}} G
\end{equation}
so that these transformations act as diffeomorphisms on the physical fields
on-shell, i.e. their action on physical fields differs from that of the
diffeomorphisms (\ref{diffs2}) by an on-shell trivial transformation. The
transformations of $P, Q'$ are
\begin{equation}
  \delta P = \frac{1}{2} i_{\hat{\xi}} (F - G), \quad \delta Q' = -
  \frac{1}{2} (1 + \bar{\ast}) di_{\hat{\xi}} (F - G) .
\end{equation}

\subsection{Gauge Algebra}

Acting on $g, d P$ or $Q$, the gauge transformations (\ref{zsymm}),(\ref{gha}) have the gauge algebra
\begin{equation}
  [\delta_{\zeta_1}, \delta_{\zeta_2}] \approx \delta_{\zeta_{12}}
\end{equation}
where
\begin{equation}
  \zeta_{12} = [\zeta_1, \zeta_2]
\end{equation}
is the Lie bracket of the vector fields $\zeta_{1,} \zeta_2$. Thus the
on-shell gauge algebra is that of diffeomorphisms. Off-shell,  the
commutator $[\delta_{\zeta_1}, \delta_{\zeta_2}] $ acting on $d P$ or $Q$ has
extra terms involving $d F$, which vanish on-shell as a result of the field
equations (\ref{feq1}),(\ref{feq2}). When acting on $P$, the commutator $[\delta_{\zeta_1},
\delta_{\zeta_2}] P$ gives $\delta_{\zeta_{12}} P$ plus terms involving $d F$
that vanish on-shell together with a gauge transformation of the form (\ref{Pgag}). It is expected that the terms involving $dF$ can be thought of as constituting  an on-shell trivial symmetry; see e.g.\ \cite{Mamade:2025jbs}.

The algebra of the $\hat{\zeta}$ transformations again gives the
diffeomorphism algebra on-shell
\begin{equation}
  [\delta_{\hat{\zeta}_1}, \delta_{\hat{\zeta}_2}] \approx
  \delta_{\hat{\zeta}_{12}}
\end{equation}
with $\hat{\zeta}_{12} = [\hat{\zeta}_1, \hat{\zeta}_2]$ when acting on $d P,
Q$, while when acting on $P$ it gives this plus a gauge transformation (\ref{Pgag}).

Finally, the $\zeta$ and $\hat{\zeta}$ transformations commute:
\begin{equation}
  [\delta_{\zeta }, \delta_{\hat{\zeta} }] = 0.
\end{equation}

\subsection{Conserved Currents}

For the bi-metric action (\ref{2act}), it was shown in \cite{Hull:2025bqo} that the current
\begin{equation}
  \Theta_{\mu \nu} = - \frac{2}{\sqrt{- g}} \frac{\delta S}{\delta g^{\mu
  \nu}} \label{thmn}
\end{equation}
(with $g = \det \{ g_{\mu \nu} \}$) is given by
\begin{equation}
  \Theta_{\mu \nu} = \frac{1}{4(q - 1) !} g^{\rho_1 \lambda_1} \ldots
  g^{\rho_{q - 1} \lambda_{q - 1}} F_{\mu \rho_1 \ldots \rho_{q - 1}} F_{\nu
  \lambda_1 \ldots \lambda_{q - 1}} \label{thmnis}
\end{equation}
and is the conserved current corresponding to the symmetry (\ref{zsymm}). This
current reduces to (\ref{Tmn}) in the linearised theory. The current
\begin{equation}
  \bar{\Theta}_{\mu \nu} = - \frac{2}{\sqrt{- \bar{g}}} \frac{\delta S}{\delta
  \bar{g}^{\mu \nu}}
\end{equation}
is  
\begin{equation}
  \bar{\Theta}_{\mu \nu} = \frac{1}{4(q - 1) !} \bar{g}^{\rho_1 \lambda_1}
  \ldots \bar{g}^{\rho_{q - 1} \lambda_{q - 1}} (G_0)_{\mu \rho_1 \ldots
  \rho_{q - 1}}  (G_0)_{\nu \lambda_1 \ldots \lambda_{q - 1}}
\end{equation}
and agrees with (\ref{Tmnh}). This is the conserved current for the symmetry
(\ref{chsym}).

For the tri-metric action (\ref{actnew}), the current (\ref{thmn}) is again
given by (\ref{thmnis}) and is the conserved current for the symmetry
(\ref{zsymm}),(\ref{gha}). The current
\begin{equation}
  \hat{\Theta}_{\mu \nu} = - \frac{2}{\sqrt{- \hat{g}}} \frac{\delta S}{\delta
  \hat{g}^{\mu \nu}}
\end{equation}
is given by
\begin{equation}
  \hat{\Theta}_{\mu \nu} = \frac{1}{4(q - 1) !} \hat{g}^{\rho_1 \lambda_1}
  \ldots \hat{g}^{\rho_{q - 1} \lambda_{q - 1}} G_{\mu \rho_1 \ldots \rho_{q -
  1}} G_{\nu \lambda_1 \ldots \lambda_{q - 1}}
\end{equation}
and is the conserved current for the symmetry (\ref{chit}),(\ref{chitp}).

\section{Discussion}

The action for self-dual gauge fields in $4 n + 2$ dimensions is
\begin{equation}
  S = \int \{ Q' \wedge dP + V (Q' - \Omega) - \hat{V} (Q' + \Omega) \}
  \label{finalact}
\end{equation}
with $\Omega$ given by (\ref{omis}). This involves novel couplings to three
metrics, $g, \hat{g}, \bar{g}$ which are not the familiar geometric couplings
but which can be understood in terms of the geometry developed in \cite{Hull:2023dgp}. The
action has the three spin-two gauge symmetries discussed in subsection \ref{trisect}
corresponding to the three spin-two gauge fields $g, \hat{g}, \bar{g}$. The
field equations for the field strengths $F, G$ are independent of the
background metric $\bar{g}$, but the action depends explicitly on it. The
action of \cite{Hull:2023dgp} is recovered by setting $\hat{g} = \bar{g}$ (or equivalently
taking $\hat{\kappa} = 0$) while Sen's action \cite{Sen:2015nph,Sen:2019qit} is recovered by setting
$\hat{g} = \bar{g} = \eta$. The tri-metric formulation is remarkably symmetric
between the physical and shadow sectors, as can be seen from the transformations
(\ref{z2sym}),(\ref{z2symf}) that preserve the field equations while changing the sign of the action.

The IIB superstring field theory of \cite{Hull:2025mtb} gives an action for the self-dual RR
gauge fields that agrees with (\ref{finalact}) to cubic order. It was shown in
\cite{Hull:2025mtb} that the action for the two gravitons $g, \hat{g}$ in the NS-NS sector
agrees to cubic order with
\begin{equation}
  S_{grav} = \frac{1}{\kappa^2} \int d^{10} x \sqrt{- g} R -
  \frac{1}{\hat{\kappa}^2} \int d^{10} x \sqrt{- \hat{g}} \hat{R} \label{ehil}
\end{equation}
where $R, \hat{R}$ are the Ricci scalars for the metrics $g, \hat{g}$. The
action given by the sum of (\ref{finalact}) and (\ref{ehil}) then gives a
non-linear completion of this which is fully invariant under the symmetries of
subsection \ref{trisect} and so is to be expected to be part of the low-energy effective
action for the IIB superstring arising from the SFT. This discussion helps clarify
the sense in which the theory is independent of the background metric
$\bar{g}$. Note, however, that the relation of this low-energy effective
action to the full action for these fields arising directly from the SFT is
subtle and not yet completely understood; see \cite{Hull:2025mtb,Mamade:2025jbs,Mamade:2025kfk,Mazel:2025fxj}. 

 \bigskip\bigskip
\noindent{\bf\Large Acknowledgements}:
\bigskip

\noindent 
 This research   was supported by   the STFC Consolidated Grant    ST/X000575/1. I would like to thank Neil Lambert  for useful discussions.

\end{document}